\documentstyle[12pt,aaspp4]{article}

\begin{document}

\title{PHOTON SPLITTING IN THE SUPERSTRONG MAGNETIC FIELDS OF PULSARS}

\author{Vladimir V.~Usov}

\affil{Department of Condensed Matter Physics, Weizmann Institute,
Rehovot 76100, Israel}


\begin{abstract}

We discuss the polarization selection rules for splitting of the 
two principal electromagnetic modes that propagate in a vacuum 
polarized by a superstrong magnetic field
($B>0.1B_{\rm cr}\simeq 4\times 10^{12}$~G). We show that 
below the threshold of free pair creation the selection rules found by 
Adler in the limit of weak dispersion remain unaffected by 
taking the resonant effects into consideration, i.e., splitting of  
one mode is strictly forbidden, while splitting of the other
is allowed.


\end{abstract}

\newpage

\section{Introduction}
Current models for the generation of coherent radio emission of pulsars
require formation of free $e^+e^-$ pairs in the pulsar magnetospheres
(e.g., Ruderman \& Sutherland \cite{RS75}; Michel \cite{M91};
Melrose \cite{M93}; Usov \cite{U96}). Such pairs may be created 
via conversion of $\gamma$-rays into $e^+e^-$ pairs in a strong 
magnetic field, $\gamma + B\rightarrow e^+ +e^- +B$ (Erber \cite{E66}). 
Besides, some (if not all) pulsars are powerful sources of $\gamma$-ray 
emission (e.g., Hartmann \cite{H95}; Thompson et al. \cite{T95};
Daugherty \& Harding \cite{DH96}). Therefore,  
propagation of $\gamma$-rays in the magnetosphere is one of
the central problems in the theory of pulsars (Usov \& Melrose
\cite{UM95,UM96}; Harding, Baring \& Gonthier \cite{HBG97}; 
Baring \& Harding \cite{BH01} and references therein).

In the magnetospheres of pulsars the plasma dispersion for $\gamma$-rays
is negligible, and to understand the process of $\gamma$-ray 
propagation in the vicinity of a pulsar it suffices to consider
propagation in the vacuum polarized by a strong magnetic field
(e.g. Usov \& Melrose \cite{UM95}; Bulik \cite{B98}). The 
principal modes of propagation for a photon in the magnetized vacuum
are linearly polarized with the photon electric field either
perpendicular ($\perp$ mode) or parallel ($\parallel$ mode) to the 
plane formed by the external magnetic field ${\bf B}$ and the photon 
wave vector ${\bf K}$. Both modes are generated in the magnetospheres 
of pulsars (Baring \& Harding \cite{BH01}). 
A single photon in either mode can decay into 
a free $e^+e^-$ pair provided that a relevant threshold, $\varepsilon
_\gamma\sin \vartheta =2mc^2$ for $\parallel$ mode and $\varepsilon_
\gamma\sin \vartheta =mc^2\eta$ for
$\perp$ mode, is exceeded, where $\varepsilon_\gamma$ is the photon
energy, $B_{\rm cr}=m^2c^3/e\hbar =4.4\times 10^{13}$~G, 
$\eta=1+[1+(2B/B_{\rm cr})]^{1/2}$, and
$\vartheta$ is the angle between the photon wave vector ${\bf K}$
and the magnetic field ${\bf B}$. For typical magnetic fields of
pulsars, $B\sim 10^{12}$~G, the absorption coefficient for 
photo-pair production is at least a few orders of magnitude higher 
than the absorption coefficient for any other inelastic interaction 
of photons and
magnetic fields (Adler \cite{A71}). However, for all known pulsars, 
$\gamma$-rays generated near the neutron star surface are produced
in a state below the pair creation threshold (Usov \& Melrose
\cite{UM95}; Baring \& Harding \cite{BH01}). In this case, the process 
of pair creation by single photons is kinematically forbidden, and
the main (inelastic) interaction of photons and the magnetic 
field is a splitting into two photons
$\gamma + B\rightarrow \gamma_1 +\gamma_2
+B$ (Adler \cite{A71}; Stoneham \cite{S79}; Melrose \cite{M83};
Baring \cite{B91}). Adler (\cite{A71}) showed that in the limit of weak 
dispersion, where the refractive indices for both modes are very close 
to unity, only $\perp$ mode may undergo the splitting process.
Near the threshold of
pair creation, however, the dispersion law differs considerably from
the vacuum case, $|{\bf K}|=\omega/c$, and the weak-dispersion limit
is no longer applicable (Shabad \cite{S75}). The polarization selection
rules for photon splitting with taking into account the resonant
dispersion law near the threshold of free pair creation were discussed
by Usov \& Shabad (\cite{US83}). Later, it was shown that bound
$e^+e^-$ pairs (positronium) may be created by single photons in a 
strong magnetic field (Shabad \& Usov \cite{SU85}, \cite{SU86};
Herold, Ruder, \& Wunner \cite{HRW85}; Usov \& Shabad \cite{US85};
Usov \& Melrose \cite{UM95,UM96}). The threshold of bound pair creation 
is slightly below the threshold of free pair creation, and therefore,
creation of bound pairs may be allowed while creation of free pairs
is forbidden yet. The dispersion curves corrected by 
taking into account formation of the mixed photon-positronium 
states were calculated in (Shabad \& Usov 
\cite{SU86}). These dispersion curves near and below the thresholds 
of pair creation differ significantly from the photon 
dispersion curves calculated by Shabad (\cite{S75}) and
used in (Usov \& Shabad \cite{US83}). In this paper, using 
the corrected dispersion curves (Shabad \& Usov \cite{SU86}) the 
polarization selection rules for splitting of both modes in a 
superstrong ($B>0.1B_{\rm cr}$) magnetic field are considered. 
 
\section{The polarization selection rules for splitting}

In a time-independent, spatially uniform external magnetic field 
${\bf B}$, the dispersion curve for the principal mode
of polarization $a$ 
$(a=\{\parallel, \perp\})$ may be
written in the following general form (e.g., Shabad \cite{S75};
Usov \& Shabad \cite{US83})

\begin{equation}
(\omega/c)^2 - K^2_\parallel = f_a(K^2_\perp )\,,
\end{equation}

\noindent
where $\omega= \varepsilon_\gamma/\hbar$ is the frequency
of state (photon, positronium atom or their mixture),
$K_\parallel$ and $K_\perp$ are the components of the 
wave vector along and across the magnetic field ${\bf B}$,  
and $f_a(K_\perp^2 )$ is a certain function of $K^2_\perp$. 
In a superstrong magnetic field for the $\parallel$ mode 
below the threshold of free pair creation, 
$(\omega /c)^2- K_\parallel^2 <(2mc^2/\hbar )^2$, we have 
(Usov \& Shabad \cite{US85};
Shabad \& Usov \cite{SU85,SU86})

\begin{equation}
f_\parallel (K^2_\perp )
={1\over 2}\left\{\left({\varepsilon_{00}(0,K^2_\perp )
\over c \hbar}\right)^2 + K^2_\perp
-\sqrt{\left[\left({\varepsilon_{00}(0,K^2_\perp )
\over c\hbar}\right)^2-K^2_\perp
\right]^2 + 4 A(K^2_\perp )}
\,\right\}\,,
\end{equation}

\noindent
where 

\begin{equation}
A(K^2_\perp )={4\alpha eB\over c^2\hbar^2}
\varepsilon_{00}(0, K^2_\perp )|\psi(0)|^2
\exp \left(-{c\hbar K^2_\perp\over 2eB}\right),
\end{equation}

\noindent
$\varepsilon_{00}(0,K^2_\perp )=2mc^2- \Delta\varepsilon_{00}
(0, K_\perp ^2)$ is the rest energy of positronium in the ground state 
when the Coulomb quantum number and the Landau numbers of the electron 
and the positron are zero ($n_c=n_-=n_+=0$),

\begin{equation}
\Delta\varepsilon_{00}(0, K_\perp^2)=\alpha^2mc^2
\left[\ln {a\over r_{_{\rm B}}\sqrt{1+r^2_{_{\rm B}}K^2_\perp}}
\right]^2
\end{equation}

\noindent
is the binding energy, $\psi (0)$ is the longitudinal wave function 
of the positronium with the constituent particles taken at the same 
points ($z^+=z^-$), 

\begin{equation}
|\psi (0)|^2={1\over a}\ln {a\over r_{_{\rm B}}\sqrt{1+r^2_{_{\rm B}}
K^2_\perp}}\,, 
\end{equation}

\noindent
$a=\hbar /me^2$ is the Bohr radius, $r_{_{\rm B}}=
(c\hbar /eB)^{1/2}$ is the radius of the electron orbit, and
$\alpha=e^2/c\hbar$ is the fine structure constant.
The dispersion curve of the $\parallel$ mode which is given by
equations (1)-(5) is photon-like for $K^2_\perp < (2mc/\hbar )^2$, 
and positronium-like for $K^2_\perp > (2mc/\hbar )^2$ (see Fig.~1). 
At $K^2_\perp\simeq (2mc/\hbar )^2$, this curve describes the 
mixed photon-positronium state. 

The positronium states of the lowest series ($n_+=n_-=0$ and $n_c\geq
0$), including the boundary of its continuum $(\omega/c)^2 -K_
\parallel^2=(2mc/\hbar )^2$, do not contribute into the polarization 
operator of a $\perp$-polarized state (Shabad \& Usov \cite{SU86}). 
Therefore, the dispersion curve $(\omega/c)^2-K^2_\parallel=
f_\perp (K^2_\perp )$ for the $\perp$ mode crosses 
the positronium spectra of the lowest series without interfering
with them (see Fig.~1). As $K^2_\perp $ grows further, this curve 
approaches eventually the dispersion curve of positronium at the 
state with $n_+ +n_-=1$ and $n_c=0$ when either the electron 
or the positron is at the first excited Landau level (Shabad 
\& Usov \cite{SU86}). 

At $B>0.1B_{\rm cr}$, the geometrical optics or adiabatic 
approximation is valid for propagation of $\gamma$-rays in the 
pulsar magnetospheres (Usov \& Melrose \cite{UM95} and references 
therein). In this case, the curvature of the magnetic field lines 
causes $\gamma$-rays of both modes to shift along the relevant
dispersion curves (see Fig.~1). In the process of propagation, 
$\gamma$-rays that are created significantly below the threshold of 
pair creation can pass through the mixed photon-positronium 
states and turns into positronium at $| K_\perp |> 2mc/\hbar$.
In this section, we discuss the polarization selection rules 
for splitting of both polarization modes irrespective of that the
the initial state is a photon or a positronium atom. We consider 
the case when the initial state is below the threshold of free
pair creation, $(\omega/c)^2-K^2_\parallel < (2mc^2/\hbar )^2$.

The splitting reactions that may be in principle in a superstrong
magnetic field are $\parallel\rightarrow \parallel 
+\parallel$, $\parallel\rightarrow\parallel + \perp$,
$\parallel\rightarrow\perp + \parallel$, $\parallel\rightarrow
\perp +\perp$, $\perp\rightarrow\perp +
\perp$, $\perp\rightarrow\perp +\parallel$, $\perp\rightarrow
\parallel + \perp$ and $\perp\rightarrow \parallel + \parallel$.  
At first we consider the kinematics of the splitting reaction
$\parallel\rightarrow \parallel +\parallel$ where the initial and 
final states are the $\parallel$ mode. We use the frame where 
$K_\parallel=0$ for the initial state. If the 
energy of the initial state is $\hbar\omega<\varepsilon_{00}(0, 
4m^2c^2)$ this reaction is the photon splitting $\gamma + 
B\rightarrow\gamma_1 +\gamma_2 +B$, while at $\varepsilon_{00}(0, 
4m^2c^2)<\hbar \omega <2mc^2$ it is the positronium annihilation 
$(e^+,e^-) +B\rightarrow \gamma_1 +\gamma_2 +B$ (see Fig.~1). 
Below, both these processes are called splitting into photons.

The splitting reaction into two photons
is kinematically allowed only if 
energy-momentum conservation can be satisfied:

\begin{equation}
\omega=\omega_1+\omega_2\,\,\,\,\,\,{\rm and} \,\,\,\,\,\,
{\bf K} ={\bf K}_1+
{\bf K}_2\,,
\end{equation}

\noindent
In our case, when $K_\parallel =0$, we have $K_
{1,\parallel}= - K_{2,\parallel}$.

Using equations (2) - (5),
we can solve the dispersion relation $(\omega /c)^2
-K^2_\parallel=f_\parallel (K^2_\perp )$ and find $|K_\perp |$ 
as a function of $\tilde\omega =\sqrt{\omega^2-c^2K^2_\parallel}$,
$|K_\perp |= \varphi_\parallel (\tilde\omega )$, where $\tilde\omega$
is the frequency in the frame where $K_\parallel$ is equal to zero. 
The function $\varphi_\parallel (\tilde\omega )$ 
is equal to zero at $\tilde\omega =0$ and increases 
monotonously with increase of $\tilde\omega$ (Shabad \& Usov 
\cite{SU86}). From equations (2)~-~(5), it follows that 
the deviation of $|K_\perp |$ from the vacuum dispersion curve
$|K_\perp |=(\tilde\omega/c)$ to the side of large
values of $|K_\perp |$ increases with increase of 
$\tilde\omega$ throughout the interval $0<\tilde\omega<2mc^2/\hbar$
(see Fig.1). In other words, we have 
$\varphi_\parallel (\tilde\omega_1)> 
\varphi_\parallel(\tilde\omega_2)$ for $\tilde\omega_1 
>\tilde\omega_2$ and $\varphi_\parallel (\tilde\omega_1 + \tilde
\omega_2 )>\varphi_\parallel (\tilde\omega_1)+\varphi_\parallel 
(\tilde\omega_2)$ for any values of $\tilde\omega_1$ and 
$\tilde\omega_2$ at $\tilde\omega_1 +\tilde\omega_2<2mc^2/\hbar$.
These properties of $\varphi_\parallel(\tilde\omega)$ and equation 
(6) yield

$$
|K_\perp| =\varphi_\parallel(\omega)=
\varphi_\parallel (\omega_1 +\omega_2)
>\varphi_\parallel (\omega_1) +\varphi_\parallel (\omega_2)
$$

$$
\geq\varphi_\parallel (\sqrt{\omega_1^2- c^2K^2_{1,\perp}})
+ \varphi_\parallel (\sqrt{\omega_2^2- c^2K^2_{2,\perp}})
$$

\begin{equation}
=\varphi_\parallel (\tilde\omega_1 )+\varphi_\parallel
(\tilde\omega_2)
=|K_{1,\perp}|+|K_{2,\perp}|\,.
\end{equation}

Therefore, the conservation of $K_\perp$ cannot be satisfied 
for the splitting reaction $\parallel\rightarrow \parallel +
\parallel$ below the threshold of free pair creation,
and this reaction is kinematically forbidden.
This is valid irrespective of that the initial state 
is either a photon or a positronium atom. 

The splitting reactions $\parallel\rightarrow \parallel +\perp$,
$\parallel\rightarrow \perp +\parallel$ and $\parallel\rightarrow
\perp +\perp$ are also kinematically forbidden 
below the threshold of free pair creation. Indeed, if we 
rewrite the dispersion relation $(\omega/c)^2-K^2_\parallel=
f_\perp (K^2_\perp )$ for the $\perp$ mode
in the form $|K_\perp |=\varphi_\perp (\tilde\omega)$,
the inequality $\varphi_\parallel (\tilde\omega)>\varphi_\perp
(\tilde\omega )$ holds throughout the interval $0<\tilde\omega 
<2mc^2/\hbar$ (Shabad \cite{S72,S75}; Shabad \& Usov \cite{SU86}). 
Using this inequality 
the chain of inequalities (7) may be continued for splitting
reactions when the initial state is the $\parallel$ mode and
at least one final state is the $\perp$ mode.
Finally, we have $|K_\perp|> |K_{1,\perp}|+ |K_{2,\perp}|$ for 
all these reactions. Hence, splitting of the $\parallel$ mode 
is strictly forbidden below the threshold of free pair creation,
i.e., Adler's conclusion remains unaffected by taking
the resonant effects into consideration. 

Since the dispertion curve $|K_\perp |=\varphi_\perp 
(\tilde\omega )$ for the $\perp$ mode crosses the positronium 
spectra of the lowest series $(n_+=n_-=0$ and $n_c\geq 0)$ 
without interfering with them, the resonant effects are 
unessential for splitting of the $\perp$ mode
below the threshold of free pair creation, $\tilde\omega 
\leq 2mc^2/\hbar$ (see Fig.~1). In this case, the splitting 
reaction $\perp\rightarrow \perp +\perp$ is kinematically 
forbidden while the splitting reactions $\perp\rightarrow 
\parallel + \parallel$, $\perp\rightarrow \parallel +\perp$ and 
$\perp\rightarrow \perp +\parallel$ are kinematically allowed
(Adler \cite{A71}). Anyone can also come to this conclusion
from the following properties of the functions $\varphi_\perp 
(\tilde\omega )$ and $\varphi_\parallel (\tilde\omega )$
below the threshold of free pair creation,
$\varphi_\perp (\tilde\omega_1)> \varphi_\perp (\tilde\omega_2)$ 
for $\tilde\omega_1>\tilde\omega_2$,  $\varphi_\perp 
(\tilde\omega_1+\tilde\omega_2 )>\varphi_\perp 
(\tilde\omega_1)+\varphi_\perp (\tilde\omega_2)$, and $\varphi_
\parallel (\tilde\omega )>\varphi_\perp (\tilde\omega )$
(Shabad \& Usov \cite{SU86}). The splitting reactions 
$\perp\rightarrow \parallel +\perp$ and $\perp\rightarrow
\perp +\parallel$ that involve an odd number of 
$\parallel$-polarized photons are forbidden by CP invariance 
in the limit of zero dispersion (Adler \cite{A71}). In the
magnetized vacuum, dispersive effects guarantee a nonzero
probability for these reactions. Therefore, the channels 
$\perp\rightarrow \parallel +\perp$ and $\perp\rightarrow
\perp +\parallel$ are strongly suppressed in comparison with
$\perp\rightarrow \parallel + \parallel$ when the refraction index 
of the magnetized vacuum is close to unity, $n-1\sim 0.1\alpha
(B/B_{\rm cr})^2\ll 1$ (Adler \cite{A71}). However, this 
suppression may not hold in an extremely strong magnetic 
fields $(B\gtrsim (10/\alpha )^{1/2}B_{\rm cr}\sim 10^{15}$~G).

\section{Discussion}
In this Letter, we have discussed splitting of the principal
modes that propagate in the vacuum polarized by a superstrong
($B>0.1B_{\rm cr}$) magnetic field. We have
shown that splitting of the $\parallel$ mode is
strictly forbidden below the threshold of free pair creation. 
This is valid irrespective of that the initial state 
is either a photon ($|K_\perp| <2mc/\hbar$)
or a positronium atom ($|K_\perp| >2mc/\hbar$).  
Splitting of the $\perp$ mode is allowed, and only the one channel 
($\perp\rightarrow\parallel +\parallel$) of splitting operates if 
the magnetic field is not extremely high (Adler \cite{A71}).

In the study of the polarization selection rules for splitting, Adler 
(\cite{A71}) has used the refractive indices

\begin{equation}
n_{\parallel ,\perp}= {c|{\bf K}|\over\omega }=
\sqrt{{\varphi_{\parallel , \perp}^2(\tilde\omega ) +K^2_\parallel}
\over{(\tilde\omega/c)^2+K^2_\parallel }}
\end{equation} 

\noindent
while we use the dispersion curves $|K_\perp |=\varphi_{\parallel,
\perp}(\tilde\omega )$. From equation (8), the properties of 
$n_{\parallel,\perp}$ used by Adler (\cite{A71}) follow immediately
from the properties of $\varphi_{\parallel,\perp}$ used in our 
study, and therefore, both these methods are equivalent.

The fact that the resonant effects do not change the polarization 
selection rules found by Adler
for photon splitting in the limit of weak dispersion 
may be easily explained. Indeed, to formulate the selection 
rules for the case when photons are significantly below the threshold 
of pair creation the following properties of functions
$\varphi_\parallel (\tilde\omega )$ and $\varphi_\perp 
(\tilde\omega )$ has been used, in fact, by Adler (\cite{A71}),
$\varphi_\parallel (\tilde\omega ) >\varphi_\perp (\tilde\omega )$
and for both principal modes the deviation of $|K_\perp |$ from the 
vacuum dispersion curve $|K_\perp |=(\tilde\omega/c)$ to the side of 
large values of $|K_\perp |$ increases with increase of $\tilde
\omega$. These properties only strengthen by the resonant 
effects, especially when the dispertion curve of the $\parallel$
mode approaches the positronium state at $|K_\perp |\gg 2mc/\hbar$
(see Fig.~1). Therefore, for a positronium atom at the
ground state ($n_+=n_-=n_c=0$) with $|K_\perp |\gg 2mc/\hbar$ 
its stability against spontaneous decay into photons 
is evident, and it has been mentioned long ago (Herold et al. 
\cite{HRW85}).

The dispersion curves calculated in (Shabad \& Usov \cite{SU86}) 
and used in this Letter
are exact relativistic consequences of the adiabatic approximation, 
and therefore, they are valid for an arbitrary strong magnetic 
field. [This relates to equations (1), (2) and (3).] 
The nonrelativity is introduced
only when the binding energy of the positronium (4) and the 
longitudinal wave function (5) that are the solutions of the 
nonrelativistic Schr\" odinger equation are substituted into
equation (3). The 
relativistic corrections to both the binding energy and the 
longitudinal wave function are small if the positron
and the electron remain nonrelativistic along the field
direction, i.e., if $\Delta \varepsilon_{00}(0,0)\ll mc^2$
(Angelie \& Deutsch \cite{AD78}; Lai \& Salpeter
\cite{LS95}). From equation (4), it follows that 
the dispersion curves calculated in (Shabad \& Usov \cite{SU86})
and therefore, our consideration are valid even if the magnetic field
is as high as the virial magnetic field ($\sim 10^{17}-10^{18}$ G) for 
neutron stars.

There is now compelling evidence for the existence of "magnetars" -
neutron stars with magnetic field strenghths in excess of $B_{\rm cr}$.
The evidence comes primarily from
observations of rapid spin down in the pulsations observed from soft 
$\gamma$-ray repeaters (e.g., Kouveliotou et al. \cite{K98}, \cite{K99}).
Recently, radio surveys have discovered a few pulsars with the surface 
magnetic fields approaching $10^{14}$ (Camilo et al. 2000).
Kulkarni \& Frail (\cite{KF93}) and Kouveliotou et al. (\cite{K94})
estimate that the birthrate of magnetars is about 10\% that of ordinary 
pulsars. The processes of photon splitting and bound pair creation
may be very important for magnetars (Thompson \& Duncan \cite{TD95};
Usov \& Melrose \cite{UM95}; Baring \& Harding \cite{BH01} and references
therein). We briefly discuss these processes in the 
magnetospheres of pulsars with very strong magnetic fields.

For pulsars with the surface magnetic field $B_{\rm S}\gtrsim 10^{13}$
G, $\perp$-polarized $\gamma$-rays generated near the 
neutron star surface split into $\parallel$-polarized 
$\gamma$-rays before they reach 
the pair creation threshold (e.g., Baring \& Harding \cite{BH01}). 
In turn, in such a strong magnetic field
$\parallel$-polarized $\gamma$-rays create bound 
pairs which are stable, in the absence 
of such external factors as electric fields and ionizing radiation
(Usov \& Melrose \cite{UM95} and references therein).
The bound pairs form a gas of electroneutral particles.
Such a gas does not undergo plasma processes, like plasma 
instabilities, which are responsible for the generation of coherent
radio emission of pulsars. Maybe, the suppression of free pair
creation in superstrong magnetic fields results in a death line of 
pulsars at $B_{\rm S}\sim B_{\rm cr}$ (e.g., Baring \& Harding
\cite{BH98}; Heyl \& Kulkarni \cite{HK98}). 
For this suppression it is very important 
that splitting of $\perp$-polarized $\gamma$-rays into 
$\parallel$-polarized $\gamma$-rays prevents from formation
of bound pairs at the state with $n_++n_-=1$ when either the 
electron or the positron is in an excited state with the spin 
quantum number, $s=1$, opposite to that in the ground state. 
Otherwise, bound pairs in the excited state
may be ionized by the recoil from photons 
radiated in a spin-flip transition, and 
free pairs may form in the magnetospheres of pulsars.

Since bound pairs do not screen the electric field $E_\parallel$,
the formation of bound rather than free pairs may cause
increases in 
the height of the polar gap and the total power carried
away by relativistic particles from
the polar gap into the pulsar magnetosphere by large factors
(e.g., Shabad \& Usov \cite{SU85}). For the case 
when there is no emission of particles from the surface
of a strongly magnetized neutron star,
these increases were considered 
by Usov \& Melrose (\cite{UM95}, \cite{UM96}) in detail.
Recently, the polar gap model, which is rather close 
to reality for typical pulsars, is developed for the case when 
particles flow freely from the stellar surface (Muslimov \& Harding 
\cite{MH97}; Harding \& Muslimov \cite{HM98}; Zhang \& Harding 
\cite{ZH00}; Hibschman \& Arons \cite{HA01}). In this model, 
the $E_\parallel$ field in the polar gap is due to the effect of
general relativistic frame dragging discovered by Muslimov \&
Tsygan (\cite{MT92}). For the pulsar period $P\gtrsim 0.1$ s and  
$0.1B_{\rm cr}\lesssim B_{\rm S}\lesssim 10B_{\rm cr}$,
this field (e.g., Harding \& Muslimov \cite{HM98})
is significantly smaller that the value $E^{\rm ion}_\parallel
\simeq (1-2)\times 10^{10}$ V~cm$^{-1}$, at which filed ionization 
of bound pairs becomes important (Usov \& Melrose \cite{UM96}). 
The mean free path of bound pair
photoionization is $l_{\rm ph}\sim 10^5(\Gamma/10^2)^{3}(T_{\rm S}
/10^6~{\rm K})^{-2}$ cm (Usov \& Melrose \cite{UM95}), where $\Gamma$ is
the Lorentz factor of bound pairs and $T_{\rm S}$ is the surface 
temperature. For high-energy bound pairs ($\Gamma>10^3-10^4$)
that might be responsible for the $E_\parallel$-field screening 
after their ionization, we have that $l_{\rm ph}$ is
larger than the polar gap height even if $T_{\rm S}$ is as high as
${\rm a~few}\times 10^6$ K. Therefore, the polar gap model 
developed in (Muslimov \& Harding 
\cite{MH97}; Harding \& Muslimov \cite{HM98}; Zhang \& Harding 
\cite{ZH00}; Hibschman \& Arons \cite{HA01}) is probably accurate 
for magnetic fields $B_{\rm S}\lesssim (0.1-0.2) B_{\rm cr}$
while for $B_{\rm S}>0.2B_{\rm cr}$ it is necessary to modify
this model by taking into account photon splitting
and bound pair creation. 

In the vicinity of pulsars with $B_{\rm S}\gtrsim 10^{13}$ G
splitting of $\perp$-polarized $\gamma$-rays into $\parallel$-polarized 
$\gamma$-rays provides a mechanism for the production of 
linearly polarized $\gamma$-rays. At energies $\varepsilon_\gamma
\lesssim 10^2$ MeV,
the $\gamma$-ray polarization may be up to 100\% (Shabad \& Usov
\cite{US83}). By observing the polarization of the $\gamma$-ray 
emission of pulsars it would be possible to estimate the
magnetic field near the pulsar surface.

\begin{acknowledgements}
I thank Janusz Gil for fruitful discussions and the referee for
helpful suggestions.
This work was supported in part by MINERVA Foundation, Munich, Germany.
\end{acknowledgements}

\clearpage

\begin{figure}
\plotone{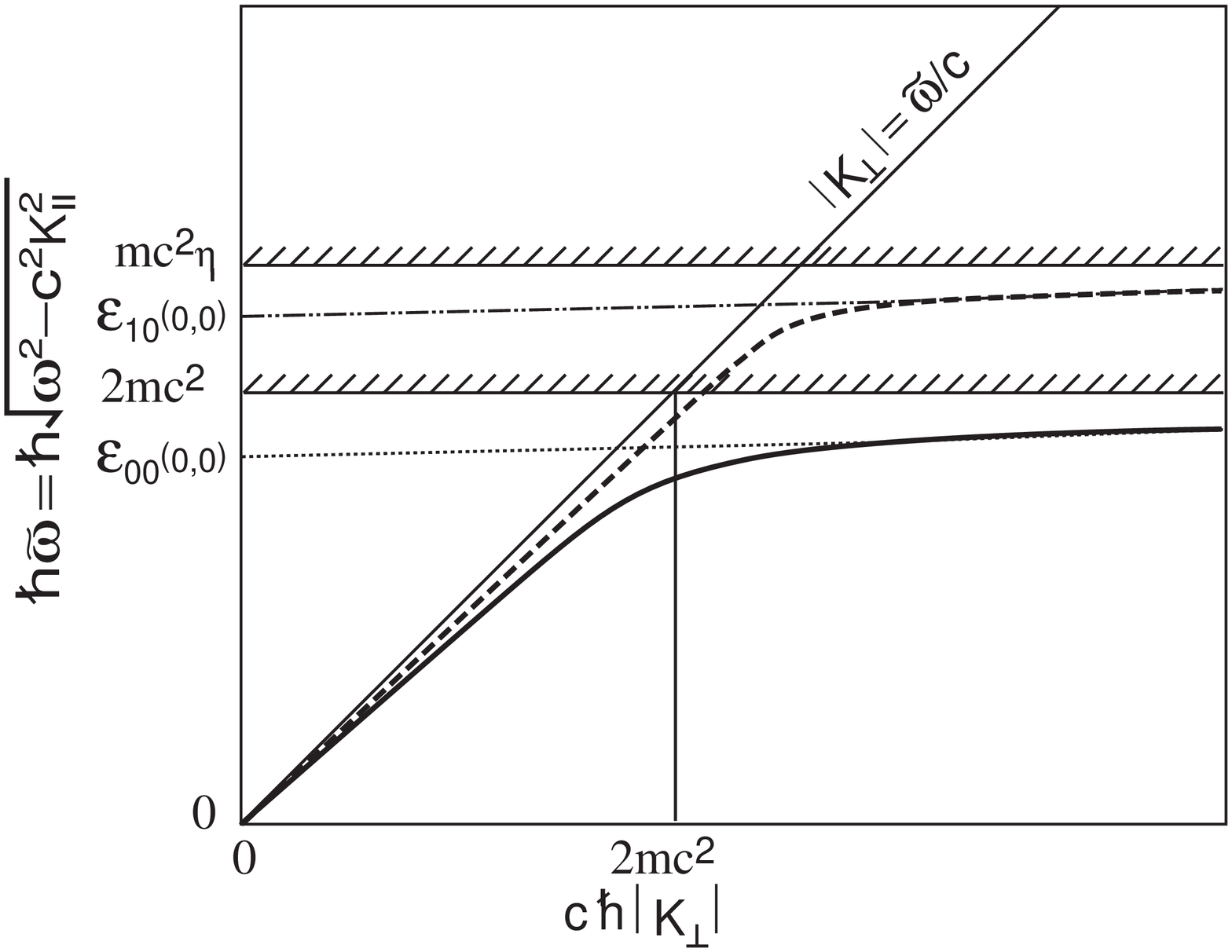}
\caption{Dispersion curves for the $\parallel$ mode (thick solid line)
and for the $\perp$ mode (thick dashed line). The sloped straight
line is the vacuum dispersion curve, $|K_\perp |=\tilde\omega/c$.
Positronium dispersion curves for the states with $n_+=n_-=n_c=0$
and with $n_++n_-=1$ and $n_c=0$ are shown by the dotted line and 
by dot-dashed line, respectively. The horizontal straight dashed 
lines are the boundaries of continua for the first two series of 
bound states with $n_+=n_-=0$ or $n_++n_-=1$. There is an infinite
number of branches for the states with $n_c\geq 1$
between the continuum boundaries and the 
positronium dispertion curves for the states with $n_c=0$
(Shabad \& Usov \cite{SU85,SU86}). They are not shown for simplicity.}
\end{figure}


\newpage

\begin{thebibliography}{}
\bibitem[1971]{A71} Adler, S.L. 1971, Ann. Phys., 67, 599

\bibitem[1978]{AD78} Angelie, C., \& Deutsch, C. 1978, Phys. Lett.,
67A, 353

\bibitem[1991]{B91} Baring, M.G. 1991, A\&A, 249, 581

\bibitem[1998]{BH98} Baring, M.G., \& Harding, A.K. 1998, ApJ,
507, L55,

\bibitem[2001]{BH01} Baring, M.G., \& Harding, A.K. 2001, ApJ,
547, 929

\bibitem[1998]{B98} Bulik, T. 1998, Acta Astron., 48, 695

\bibitem[2000]{C00} Camilo, F., et al. 2000, ApJ, 541, 367 

\bibitem[1996]{DH96} Daugherty, J.K., \& Harding, A.K. 1996,
ApJ, 458, 278

\bibitem[1966]{E66} Erber, T. 1966, Rev. Mod. Phys., 38, 626

\bibitem[1997]{HBG97} Harding, A.K., Baring, M.G., \& Gonthier,
P.L. 1997, ApJ, 476, 246

\bibitem[1998]{HM98} Harding, A.K., \& Muslimov, A.G. 1998, ApJ,
508, 328

\bibitem[1995]{H95} Hartmann, D.H. 1995, A\&A Rev., 6, 225

\bibitem[1985]{HRW85} Herold, H., Ruder, H., \& Wunner, G.
1985, Phys. Rev. Lett., 54, 1452

\bibitem[1998]{HK98} Heyl, J.S., \& Kulkarni, S.R. 1998, ApJ,
506, L61

\bibitem[2001]{HA01} Hibschman, J., \& Arons, J. 2001, ApJ,554, 624

\bibitem[1994]{K94} Kouveliotou, C. et al. 1994, Nature, 368, 125

\bibitem[1998]{K98} Kouveliotou, C. et al. 1998, Nature, 393, 235

\bibitem[1999]{K99} Kouveliotou, C. et al. 1999, ApJ, 510, L115

\bibitem[1993]{KF93} Kulkarni, S.R., \& Frail, D.A. 1993, Nature, 365, 33

\bibitem[1995]{LS95} Lai, D., \& Salpeter, E.E. 1995, Phys.
Rev. A, 52, 2611

\bibitem[1983]{M83} Melrose, D.B. 1983, Aust. J. Phys., 36, 775


\bibitem[1993]{M93} Melrose, D.B. 1993, in Pulsars as Physics
Laboratories, ed. R.D. Blandford et al. (Oxford: Oxford Univ. Press),
105

\bibitem[1991]{M91} Michel, F.C. 1991, Theory of Neutron Star 
Magnetospheres (Chicago: Univ. of Chicago Press)

\bibitem[1997]{MH97} Muslimov, A.G., \& Harding, A.K. 1997, ApJ,
485, 735

\bibitem[1992]{MT92} Muslimov, A.G., \& Tsygan, A.I. 1992,
MNRAS, 255, 61

\bibitem[1972]{S72} Shabad, A.E. 1972, Lett. Nuovo Cimento, 3, 457

\bibitem[1975]{S75} Shabad, A.E. 1975, Ann. Phys., 90, 166

\bibitem[1985]{SU85} Shabad, A.E., \& Usov V.V. 1985, Ap\&SS, 117, 309

\bibitem[1986]{SU86} Shabad, A.E., \& Usov V.V. 1986, Ap\&SS, 128, 377

\bibitem[1979]{S79} Stoneham, R.J. 1979, J. Phys. A, 12, 2187

\bibitem[1996]{U96} Usov, V.V. 1996, in IAU Colloq. 160, Pulsars: 
Problems and Progress, ed. S. Johnston, M.A. Walker, \& M. Bailes
(ASP Conf. Ser. 105) (San Francisco: ASP), 323

\bibitem[1995]{UM95} Usov, V.V., \& Melrose, D.B. 1995, Australian J. 
Phys., 48, 571

\bibitem[1996]{UM96} Usov, V.V., \& Melrose, D.B. 1996, ApJ, 464, 306

\bibitem[1983]{US83} Usov, V.V., \& Shabad, A.E. 1983, Soviet Astron. 
Lett., 9, 212

\bibitem[1985]{US85} Usov, V.V., \& Shabad, A.E. 1985, Soviet
Phys.-JETP Lett., 42, 19

\bibitem[1975]{RS75} Ruderman, M.A., \& Sutherland, P.G. 1975, ApJ,
196, 51

\bibitem[1995]{TD95} Thompson, C,, \& Duncan, R.C. 1995, MNRAS, 275, 255

\bibitem[1995]{T95} Thompson, D.J. et al. 1995, ApJS, 101, 259

\bibitem[2000]{ZH00} Zhang, B., \& Harding, A.K. 2000, ApJ, 532, 1150

\end{thebibliography}
\end{document}